# DOUBLE STOCHASTIC RESONANCE IN CONDUCTORS WITH NARROW CONDUCTION BAND

E.M. Epshtein[1], T.A. Gorshenina[2], G.M. Shmelev[2]

[1]Institute of Radio Engineering and Electronics of the Russian Academy of Sciences, 141195, Fryazino, Russia, eme253@ms.ire.rssi.ru

[2]Volgograd State Pedagogical University, 400131, Volgograd, Russia, shmelev@fizmat.vspu.ru

**Abstract**. Results are presented of studying the stochastic resonance (SR) in conductors with bcc lattice and narrow conduction band. In such materials, SR has a feature, namely, the weak-signal gain as a function of the additive and non-correlated noise level can have two maxima, i.e., double SR takes place.

The stochastic resonance (SR) can take place in nonlinear multistable systems under interaction of a weak periodic signal with noise [1]. Under SR conditions, the signal gain rises with the noise intensity, reaches maximum and then drops. It was shown [2, 3], that nonequilibrium electron gas in a quasi-two-dimensional semiconductor superlattice becomes a bistable system in high electric field, so that SR appears to be possible [3].

In this report, results are presented of SR investigation in bulk conductors with narrow conduction band and bcc lattice. Unlike [3], SR has an essential feature in this case, namely, the gain as a function of the noise level can have two maxima. Such a phenomenon may be called double SR.

We take spontaneous electric field $E_y$ transverse with respect to the current in the sample as an order parameter, while $E_x$ (and/or $E_z$) field as control parameter, the coordinate axes being directed along the crystallographic ones. The electron dispersion law in bcc lattice within the tight-coupling approximation is used. The current density **j** is found in standard way by solving the Boltzmann equation with the collision term within $\tau$-approximation ($\tau$ = const). The calculation results can be presented by means of a synergetic potential $\Phi(E_x^2, E_y^2, T) = \int j_y dE_y + \text{const}$ ($j_y = \partial\Phi/\partial E_y$).

In our case,

$$\Phi(E) = \frac{\sigma_0 C(T) E_0^2}{8}\left[\ln\left(\left(E_0^2 + E_x^2 + (E_y - E_z)^2\right)^2 - 4E_x^2(E_y - E_z)^2\right) + \ln\left(\left(E_0^2 + E_x^2 + (E_y + E_z)^2\right)^2 - 4E_x^2(E_y + E_z)^2\right) - 2\ln(E_0^4)\right] + \text{const} \quad (1)$$

Here $\sigma_0 = e^2 n \Delta d^2 \tau/\hbar^2$, $E_0 = \hbar/ed\tau$, $d$ is lattice constant, $\tau$ is momentum relaxation time, $n$ is carrier density. For nondegenerate electrons, $C(T) \approx 1.025\left(I_1(\Delta/2kT)/I_0(\Delta/2kT)\right)^2$, $I_n(x)$ is modified Bessel function, $T$ is lattice temperature, $2\Delta$ is the conduction band width, $k$ is the Boltzmann constant. If the sample is open along $y$ direction ($j_y = 0$) and $E_z = 0$, then the spontaneous field is $E_{ys} = \pm\sqrt{E_x^2 - E_0^2}$ ($|E_x| > E_0$). The system states with $E_y = E_{ys}$ are stable, so that the potential has two minima at $|E_x| > E_0$.

With the current fluctuations ($\delta j_y(t)$) taking into account, the absence condition of the current along $y$ direction is

$$\frac{dE_y}{dt} = -\frac{4\pi}{\varepsilon}\frac{\partial\Phi}{\partial E_y} - \frac{4\pi}{\varepsilon}\delta j_y(t), \quad (2)$$

where $\varepsilon$ is dielectric constant. The current correlation function for thermal (additive and non-correlated) noise [4] takes the form

$$\langle \delta j_y(t)\delta j_y(t')\rangle = (\sigma_0 C(T) E_0^2/8\alpha(T))\delta(t - t'), \quad (3)$$

where $\alpha(T) = \varepsilon V E_0^2/16\pi kT$, $V$ is volume of the system. The condition (2) is nothing but the Langevin equation. From the corresponding Fokker - Planck equation [5] we find a steady-state distribution function for the random component of $E_y$: $\overline{f} = AW^{-\alpha}$, $W = ((E_0^2 + E_x^2 + E_y^2)^2 - 4E_x^2 E_y^2)/E_0^4$, $A$ is normalization factor. At $E_x = 0$, the corresponding average value $\overline{E_y^2}$ goes over into the Nyquist formula [4] with $\alpha \gg 1$, as assumed in the present work. The average transition rate of the bistable system from one stable state to another one $r_k$ (Kramers rate) [5] is determined by the formula

$$r_k^{-1}(E_x^2, T) = \frac{4\alpha(T)}{\sigma_0 C(T) E_0^2}\int_{-\sqrt{E_x^2 - E_0^2}}^{+\sqrt{E_x^2 - E_0^2}}\frac{dE_y}{\overline{f}(E_x^2, E_y^2, \alpha(T))}\int_{-\infty}^{E_y}\overline{f}(E_x^2, E_y'^2, \alpha(T))dE_y'. \quad (4)$$

It is important that the dependence of $r_k^{-1}$ on the noise level (or temperature) is nonmonotonic one and has a maximum at some value of $T$ (see Fig. 1). Such non-monotonicity due to finite width of the conduction band leads to the double SR. In that case, the gain of a weak periodic signal [1] is

$$\eta(E_x, T) = 16\alpha^2(T) E_{ys}^4 / C^2(T) E_0^4 (1 + (\Omega^2 r_k^{-2})/4), \quad (5)$$

where $\Omega$ is the frequency of the input signal along $y$ axis. Fig. 2 illustrates the double SR (although the $\eta = \eta(E_x)$ function has only one maximum). The results obtained are valid also for a simple cubic lattice, if $x$ and $y$ axes make an angle of 45° with the lattice crystallographic axes.

Note that the double SR in a quasi-two-dimensional superlattice was studied in [6], however, different conditions were considered there.

## Figure captions

Fig. 1. $r_k = r_k(T)$ function at various parameter values: $a - V = 10^{-13}$ cm$^3$, $\Delta = 10^{-2}$ eV; $b - V = 5 \times 10^{-13}$ cm$^3$, $\Delta = 10^{-2}$ eV.

Fig. 2. $\eta = \eta(T)$ function at various parameter values: $a - V = 10^{-13}$ cm$^3$, $\Delta = 10^{-2}$ eV, $\Omega = 10^{13}$ s$^{-1}$; $b - V = 5 \times 10^{-13}$ cm$^3$, $\Delta = 10^{-2}$ eV, $\Omega = 10^{13}$ s$^{-1}$.

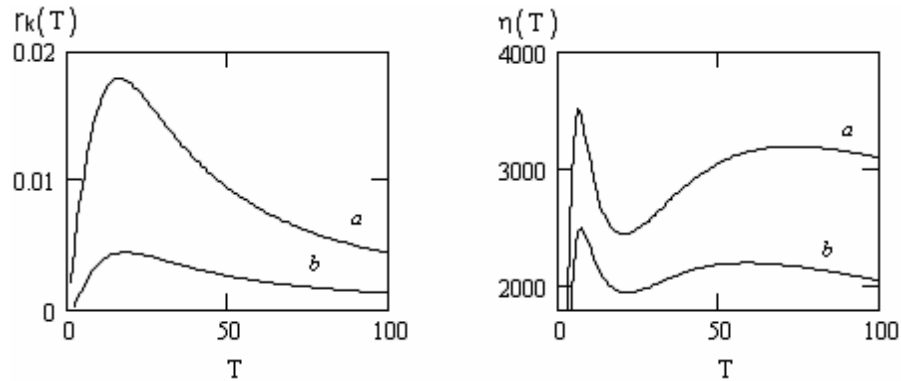